# *p*-SWAP: A Generic Cost-Effective Quantum Boolean-Phase SWAP Gate Using Two CNOT Gates and the Bloch Sphere Approach

ALI AL-BAYATY

Portland State University, USA, albayaty@pdx.edu, ORCID: 0000-0003-2719-0759

MAREK PERKOWSKI

Portland State University, USA, h8mp@pdx.edu, ORCID: 0000-0002-0358-1176

*Abstract* – A Boolean-Phase swapping gate is introduced for quantum generality and cost-effectiveness, which is termed the "*p*-SWAP gate", where *p* is a customizable phase difference for a set of swapped qubits and $0 \leq p \leq \pm \pi$ radians. The generality of the *p*-SWAP gate is proposed for quantum Phase oracles requiring a desirable *p* for a set of swapped qubits, as well as for quantum Boolean oracles when *p* is ignored. The cost-effectiveness of the *p*-SWAP gate comes from the lower quantum cost and depth for its final synthesized (transpiled) quantum circuit into a quantum computer, as compared to the standard SWAP gate. In general, the standard SWAP gate is constructed using three Feynman (CNOT) gates, while our *p*-SWAP gate only utilizes two CNOT gates. In this paper, the desirability of *p* is geometrically chosen using our proposed Bloch sphere approach, without using any matrices multiplication and unitary representations. After transpilation, the final transpiled *p*-SWAP gate has approximately 23% quantum cost reduction and 26% depth minimization than those of the final transpiled standard SWAP gate.

## 1. Introduction

In quantum computing, the quantum operation of a standard SWAP gate swaps (switches) the indices of its two input qubits into two output qubits [1–3], and these two qubits are termed the "targets". Such that, a standard SWAP gate applies on the targets $|q_1q_0\rangle \rightarrow |q_0q_1\rangle$: $|00\rangle \rightarrow |00\rangle$, $|01\rangle \rightarrow |10\rangle$, $|10\rangle \rightarrow |01\rangle$, and $|11\rangle \rightarrow |11\rangle$, where $q_0$ is the first indexed target and $q_1$ is the second indexed target. In general, the quantum circuit of a standard SWAP gate is constructed using three Feynman (CNOT) gates, where the CNOT gate is considered a cost-expensive 2-bit quantum gate. For instance, the CNOT gate requires multiple microwave pulses than a 1-bit quantum gate for IBM quantum computers [4,5]. Please observe that the "*n*-bit" is often used instead of the "*n* qubits" for a multiple-qubit gate [6–8], where $n \geq 2$.

The standard SWAP gate is an important building entity in many quantum applications and algorithms. For instance, the standard SWAP gates are mainly utilized to: (i) connect the non-neighboring physical qubits of a quantum computer [9–12], (ii) implement the quantum Fourier Transform (QFT) and the inverse QFT of Shor's algorithm and quantum phase estimation (QFE) [1–3,13,14], (iii) efficiently construct Grover's algorithm and quantum approximate optimization algorithm (QAOA) [15–18], (iv) build quantum Boolean functions and reversible circuits [19–21], (v) implement quantum cryptography and cryptarithmetic algorithms [22–24], and (vi) prepare Dicke states [25–27], just to name a few.

Subsequently, when a quantum application consisting of many standard SWAP gates is synthesized (transpiled) into a quantum computer, the final transpiled quantum circuit of such an application has a high quantum cost and depth, due to the higher utilized number of CNOT gates for those many standard SWAP gates. Please observe that the quantum cost defines the total number of the final transpiled quantum gates, which are the native "basis" gate of a quantum computer, the depth is the critical longest path along all native gates, and a higher quantum cost and depth produce a longer delay and increase the decoherence for the utilized physical qubits of a quantum computer [28–30]. For IBM quantum computers, the native gates are the Identity (I), Pauli-X (X), half-rotational X $(\sqrt{X})$, CNOT, and rotational Pauli-Z (RZ($\theta$)), where a half rotation is denoted by $+\pi/2$ radians and $0 \leq \theta \leq \pm\pi$ radians.

The goal of this paper is to introduce a generic cost-effective Boolean-Phase swapping gate, which is termed the "*p*-SWAP gate", with a customizable phase difference (*p*) for $0 \leq p \leq \pm\pi$ radians. We designed the *p*-SWAP gate based on two quantum concepts, generality and cost-effectiveness, as follows.

1. The generality of the *p*-SWAP gate comes from its utilization in both Boolean and Phase oracles [6–8] for different quantum applications and algorithms. Such that, the *p* can be ignored for the swapped qubits of a Boolean oracle or customized for a set of swapped qubits for a Phase oracle.
2. The cost-effectiveness of the *p*-SWAP gate comes from its utilization of only two CNOT gates instead of three. Such that, the quantum cost and depth of the *p*-SWAP gate are always lower than those of the standard SWAP gate, after transpilation into a quantum computer.

Our introduced *p*-SWAP gate is derived from the *i*SWAP gate proposed in [31] of two CNOT gates. However, the *p* is geometrically chosen using our proposed Bloch sphere approach, instead of performing the conventional approach of using matrices multiplication and unitary representations. The Bloch sphere approach is a "geometrical design tool" for constructing cost-effective quantum gates, based on their rotational quantum operations in the XY-plane, XZ-plane, YZ-plane, and other projectional planes along the three axes of the Bloch sphere. Collectively, the Bloch sphere and its projectional planes are termed the "Bloch sphere approach (BSA)", as discussed in the quantum BSA protocol [32] and in our two cost-effective quantum libraries: GALA-*n* (Generic Architecture of Layout-Aware *n*-bit gates) [33,34] and CALA-*n* (Clifford+T-based Architecture of Layout-Aware *n*-bit gates) [35,36], where $n \geq 2$ qubits. Both GALA-*n* and CALA-*n* quantum libraries have become part of the IBM Qiskit ecosystem [37].

For different quantum computing technologies, various research papers and studies discussed different approaches to improve the fidelities, speedups, and circuit depths of the standard SWAP gates, by re-constructing the quantum circuits of the SWAP gates using: (i) the momentum-controlled NOT (MC-NOT) and polarization-controlled NOT (PC-NOT) gates for the Silicon nanophotonics technology [38], (ii) the $\sqrt{SWAP}$ gates for the Silicon quantum dots technology [39], (iii) the $\sqrt{SWAP}$ gates for the superconducting Transmons technology [40], (iv) the Clifford+T gates for the superconducting Transmons technology [41], and (v) the cross-gate pulse cancellation and cross-resonance polarity gates for the superconducting Transmons technology [9]. However, for the aforementioned research papers and studies, we observed that the reconstructed quantum circuits of the standard SWAP gates require at least three CNOT gates. While the quantum circuit of our introduced *p*-SWAP gate only requires two CNOT gates, with a customizable *p* as well, because the purpose of our research is concentrated on introducing a cost-effective swapping gate with a lower quantum cost and circuit depth for the superconducting Transmons technology, i.e., IBM quantum computers.

In this paper, we practically prove that the *p*-SWAP gate is a generic quantum gate, by ignoring or customizing the *p* for a set of swapped qubits. After transpilation using the `ibm_brisbane` quantum computer [42,43], we experimentally prove that the *p*-SWAP gate is a cost-effective quantum gate for approximately 23% and 26% reduction of its final quantum cost and depth, respectively, as compared to the standard SWAP gate.

## 2. Methods

This section discusses how to geometrically design the *p*-SWAP gate using the Bloch sphere approach (BSA) [32], instead of utilizing the conventional design approach of unitary representations and multiplications, to generate a generic cost-effective Boolean-Phase swapping gate from two CNOT gates.

### 2.1. The Bloch Sphere

The Bloch sphere is a three-dimensional geometrical sphere of three axes (X, Y, and Z) that represents the quantum states of a qubit. When a series of quantum gates are applied to a qubit, the Bloch sphere visualizes such quantum operations in Hilbert space ($\mathcal{H}$) [1–3]. Hence, in quantum computing, the Bloch sphere is mainly utilized as a geometrical visualization (and verification) tool, and Figure 1 depicts six quantum states of a qubit, which are visualized on the three axes of the Bloch sphere.

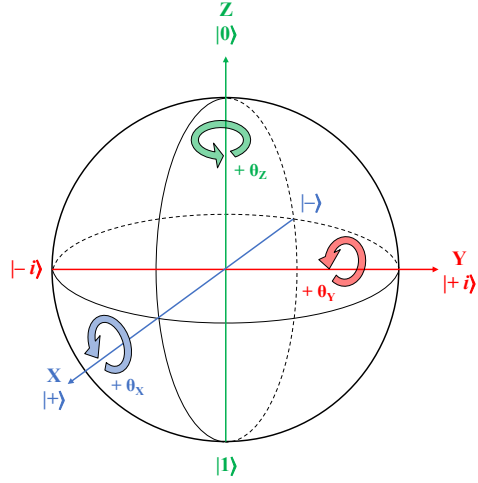

**Figure 1.** The Bloch sphere is a three-dimensional geometrical sphere consisting of three axes (X in blue, Y in red, and Z in green) and three rotational angles ($\pm\theta_X$, $\pm\theta_Y$, and $\pm\theta_Z$), where $+\theta$ represents a counterclockwise rotation and $-\theta$ represents a clockwise rotation for their respective axes [32].

A (1-bit or *n*-bit) quantum gate having X (as a Pauli-X), Y (as a Pauli-Y), or Z (as a Pauli-Z) in its notation rotates the state of a qubit around the X-, Y-, or Z-axis of the Bloch sphere, respectively, where $n \geq 2$ qubits. Such a quantum rotation around an axis of the Bloch sphere relies on a defined rotational angle ($\pm\theta$), as shown in Figure 1, where $+\theta$ represents a counterclockwise rotation and $-\theta$ represents a clockwise rotation for $0 \leq \theta \leq \pm\pi$ in radians. For instance, the following Z gates rotate the state of a qubit around the Z-axis of the Bloch sphere by defined $\theta$ angles:

- The Z gate rotates the state of a qubit by $+\pi$ radians.
- The $\sqrt[2]{Z}$ gate rotates the state of a qubit by $+\pi/2$ radians, where the $\sqrt[2]{Z}$ gate is also termed the S gate [1–3,6].
- The $\sqrt[4]{Z}$ gate rotates the state of a qubit by $+\pi/4$ radians, where the $\sqrt[4]{Z}$ gate is also termed the T gate [1–3,44].

Please observe that all (1-bit and *n*-bit) quantum gates are unitary gates [1–3,6], and a few of them are non-Hermitian gates [1–3,6], i.e., their quantum operations are not in their own inverses. For instance, the $S^\dagger$ is the inverse gate for the S gate, and the $T^\dagger$ is the inverse gate for the T gate. In addition, some rotational gates alter the phase of a qubit, such that the choice of rotational gates is a critical factor in the design of a quantum circuit. For instance, the Z gate is not the same as the IBM native RZ($+\pi$) gate, due to their global phase difference as expressed in Equation (1), where $-i$ is the global phase of the Z gate.

$$\text{RZ}(+\pi) = e^{-i\frac{\pi}{2}Z} = \begin{bmatrix} e^{-i\frac{\pi}{2}} & 0 \\ 0 & e^{i\frac{\pi}{2}} \end{bmatrix} = -iZ \tag{1}$$

### 2.2. Projectional Planes of The Bloch Sphere

In [32–37], we discussed how to geometrically design any single-target quantum gate using the BSA, i.e., the Bloch sphere and its projectional planes. The projectional planes are the two-dimensional circular planes along the three axes of the Bloch sphere. In general, different projectional planes along different axes of the Bloch sphere can be utilized to geometrically design a quantum gate, depending on (i) the quantum operational purpose of such a gate and (ii) the supported native gates of a utilized quantum computer. Hence, the BSA is a "geometrical design tool" that can visually construct any quantum gate. Figure 2 demonstrates the XY-plane, XZ-plane, YZ-plane, and *m* projectional planes (*P*) along the three axes of the Bloch sphere, where $m \geq 1$ and the black dots denote the segments of "semicircles", "quadrants", and "octants" for a projectional plane.

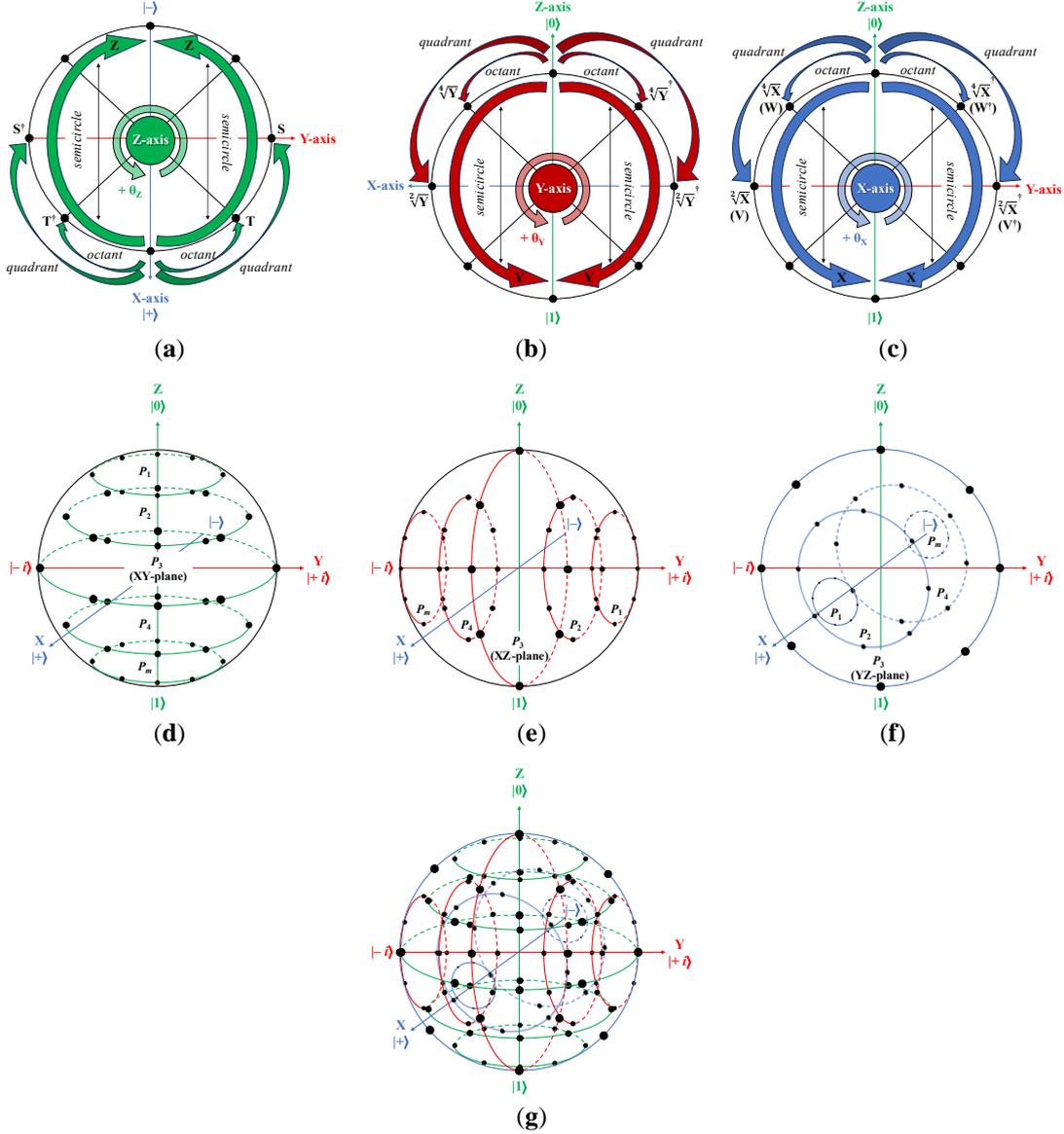

**Figure 2.** Projectional planes (*P*) along the three axes of the Bloch sphere: (**a**) the XY-plane and its segments along the Z-axis and at the center of the Bloch sphere; (**b**) the XZ-plane and its segments along the Y-axis and at the center of the Bloch sphere; (**c**) the YZ-plane and its segments along the X-axis and at the center of the Bloch sphere; (**d**) *m* projectional planes along the Z-axis of the Bloch sphere; (**e**) *m* projectional planes along the Y-axis of the Bloch sphere; (**f**) *m* projectional planes along the X-axis of the Bloch sphere; (**g**) *m* projectional planes along all three axes of the Bloch sphere, where $m \geq 1$ and the segments (as black dots) are the semicircles, quadrants, and octants for a projectional plane [32].

For IBM quantum computers, because the quantum operations of all native gates mainly rotate around the X-axis and Z-axis of the Bloch sphere, we utilize the XY-plane of the Bloch sphere (see Figure 2(a)) to visually design our generic cost-effective *p*-SWAP gate. The XY-plane is divided into a number of segments, to represent the quantum rotations ($\pm \theta$) of IBM native RZ gates applied to a qubit as follows.

1. The "semicircle" segment is half of the XY-plane that represents the quantum rotations of Z gates, i.e., RZ($\pm \pi$). Such that, the XY-plane consists of two semicircles.
2. The "quadrant" segment is one-fourth of the XY-plane that represents the quantum rotations of S and S$^\dagger$ gates, i.e., RZ($+\pi/2$) and RZ($-\pi/2$), respectively. Such that, the XY-plane consists of four quadrants.
3. The "octant" segment is one-eighth of the XY-plane that represents the quantum rotations of T and T$^\dagger$ gates, i.e., RZ($+\pi/4$) and RZ($-\pi/4$), respectively. Such that, the XY-plane consists of eight octants.

In the XY-plane, the IBM native X and √X gates rotate a qubit around the X-axis by + π and + π/2 radians, respectively. However, the IBM native CNOT gate rotates its target qubit around the X-axis by + π radians, when its control qubit is set to the $|1\rangle$ state; otherwise, there is no rotation occurs. For all IBM quantum computers of 127 qubits, the native CNOT gate has been replaced with the native ECR (echoed cross-resonance) gate [45–47] to: (i) minimize the recalibration error rates and coherent errors, and (ii) extend the duration of measurements.

## 2.3. Design of p-SWAP Gate

The quantum circuit of our generic cost-effective *p*-SWAP gate is derived from the quantum circuit of the *i*SWAP gate proposed in [31]. For that, the *p*-SWAP gate has only two CNOT gates, as compared to the standard SWAP gate of three CNOT gates shown in Figure 3(a). Therefore, the *p*-SWAP gate is considered a cost-effective quantum gate in the context of the utilized number of CNOT gates.

The *i*SWAP gate, shown in Figure 3(b), swaps its two qubits as a standard SWAP gate. However, the *i*SWAP gate only alters the phases to *i* (a phase difference of + π/2 radians), when the two input qubits are in the $|01\rangle$ and $|10\rangle$ states. In contrast, our *p*-SWAP gate includes a customizable phase difference (*p*) for a set of swapped qubits, where $0 \leq p \leq \pm \pi$ radians. Therefore, the *p*-SWAP gate is considered a generic quantum gate in the context of customizing the *p* for a set of swapped qubits.

On the one hand, the *p*-SWAP gate is a generic cost-effective Phase swapping gate for Phase oracles, since it customizes the *p*. On the other hand, when *p* of the *p*-SWAP gate is ignored, i.e., *p* can be arbitrarily set to any angular value, such a *p*-SWAP gate is a generic cost-effective Boolean swapping gate for Boolean oracles, since it consists only of two CNOT gates, as illustrated in Figure 3(c). Subsequently, the *p*-SWAP gate is a generic cost-effective Boolean-Phase swapping gate.

Please observe that our introduced *p*-SWAP gate is the generalization concept of the *i*SWAP gate proposed by the IBM quantum system [31]. However, the *i*SWAP gate has a fixed *i* phase for a fixed set of the swapped qubits in the $|01\rangle$ and $|10\rangle$ states. While the *p*-SWAP gate has customizable *p* phases for customizable swapped qubits in the $|00\rangle$, $|01\rangle$, $|10\rangle$, and/or $|11\rangle$ states.

In the quantum circuit of the *p*-SWAP gate, the customizable *p* for a set of swapped qubits is geometrically chosen using the BSA, i.e., the customizable phase-based gates of the *p*-SWAP gate are visually constructed using the BSA. Since the *p*-SWAP gate is a two-target quantum gate, two XY-planes of two Bloch spheres (one plane for each target qubit) are required to visually construct such a gate based on the desirable *p*. In this paper, the phase-based gates are chosen as IBM native RZ gates, including S, S†, and Z gates. However, the phase-based gates can be also chosen as RY gates depending on the supported native gates of a utilized quantum computer. Therefore, the *p*-SWAP gate is a generic cost-effective Boolean-Phase swapping gate on any quantum computer.

Based on that, two phase-based gates (*v* and *ω*) are utilized to generate a customizable *p* for a set of swapped qubits, as stated in Equation (2) and Equation (3) and illustrated in Figure 3(d), where A, B, C, and D are the numerical cofactors that define the final desirable *p*.

$$v = \text{RZ}\left(\frac{\pm A \cdot \pi}{B}\right) \qquad (2)$$

$$\omega = \text{RZ}\left(\frac{\pm C \cdot \pi}{D}\right) \qquad (3)$$

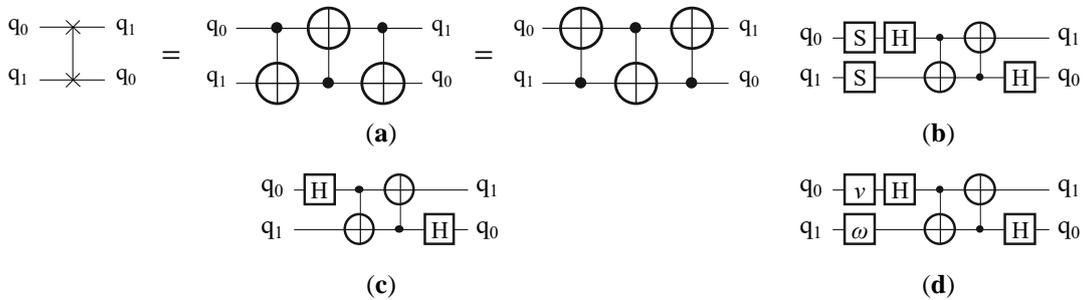

**Figure 3.** The decomposed quantum circuits of SWAP, *i*SWAP, and *p*-SWAP gates: (**a**) the standard SWAP gate; (**b**) the *i*SWAP gate [31]; (**c**) our proposed *p*-SWAP gate for Boolean applications, when *p* is ignored for the swapped qubits; (**d**) our proposed *p*-SWAP gate for Phase applications using two phase-based gates (*v* and *ω*) for a desirable *p*, as stated in Equation (2) and Equation (3), where $0 \leq p \leq \pm \pi$ radians.

Figure 4 depicts five quantum states $|\psi\rangle$ of a $p$-SWAP gate for vertical-based operational gates. Based on these $|\psi\rangle$ states, Figure 5 demonstrates the quantum transitions of the $|\psi\rangle$ states for the swapping outcomes of a $p$-SWAP gate on all permutative states of targets ($q_0$ and $q_1$), using the BSA of two Bloch spheres (two XY-planes). Note that, when $p$ is ignored, such swapping outcomes will be the same whether the $p$-SWAP gate is utilized for Boolean or Phase applications. For customizing $p$, the initially assigned $v$ and/or $\omega$ gates to the targets will guarantee the final desirable $p$ for a set of swapped qubits. For instance, in Figure 3(d), when $v = I$ and $q_0 = |1\rangle$, the first Hadamard (H) gate applied to $q_0$ will generate the superposition state of $\frac{1}{\sqrt{2}}(|0\rangle + |1\rangle)$, and no phase is acquired for $q_0$. Otherwise, when $v = Z$ and $q_0 = |1\rangle$, such an H gate will generate the superposition state of $-\frac{1}{\sqrt{2}}(|0\rangle + |1\rangle)$, and the acquired phase for $q_0$ affects the final swapping outcome.

Hence, the BSA of two Bloch spheres is utilized here as a geometrical concept to visually assign the desirable $p$ for a set of swapped qubits, without using matrix multiplications and unitary representations, and Table 1 declares the six geometrical concepts of BSA. Using Equation (1), Equation (2), and Equation (3), these six geometrical concepts state how to obtain the final desirable $p$ by assigning the required angles for $v$ and/or $\omega$ gates applied to $q_0$ and/or $q_1$, respectively.

**Table 1.** Six geometrical concepts of BSA for visually assigning the desirable $p$ for a set of swapped qubits ($q_0$ and $q_1$) using two phase-based gates ($v$ and $\omega$), where $0 \leq p \leq \pm \pi$ radians.

| Geometrical concepts | Effected set of swapped qubits $|q_1q_0\rangle \rightarrow |q_0q_1\rangle$ | Apply $v$ and $\omega$ to | Angles for $v$ and $\omega$ | Final desirable $p$ |
|---|---|---|---|---|
| 1 | Non-identical states: $|01\rangle \rightarrow |10\rangle$ $|10\rangle \rightarrow |01\rangle$ | $q_0$ and $q_1$ | Both gates have $+ \pi/2$ | $+ \pi/2$ |
| 2 | Identical states: $|00\rangle \rightarrow |00\rangle$ $|11\rangle \rightarrow |11\rangle$ | $q_0$ and $q_1$ | Both gates have $- \pi/2$ | $+ \pi/2$ |
| 3 | $|00\rangle \rightarrow |00\rangle$ | $q_0$ and $q_1$ | Both gates have $+ \pi$ | $+ \pi$ |
| 4 | $|01\rangle \rightarrow |10\rangle$ | $q_0$ | $+ \pi$ | $+ \pi$ |
| 5 | $|10\rangle \rightarrow |01\rangle$ | $q_1$ | $+ \pi$ | $+ \pi$ |
| 6 | $|11\rangle \rightarrow |11\rangle$ | None | None | $+ \pi$ |

For instance, for the first geometrical concept in Table 1, the angles of $v$ and $\omega$ gates are obtained when A = C = +1 and B = D = +2, then $v$ and $\omega$ gates are substituted by the IBM native RZ($+ \pi/2$) gates, respectively. Please observe that different desirable $p$ can be calculated for a set of swapped qubits, using different cofactors (A, B, C, and D) of Equation (2) and Equation (3). Such that, these cofactors can be further investigated for other angular values to generate various $p$-SWAP gates. Thereby, in this paper, we introduce the $p$-SWAP gate as a generic cost-effective framework for constructing various Boolean-based and Phase-based swapping gates, which fulfill the quantum demands of Boolean and Phase oracles for specific-task quantum applications and algorithms.

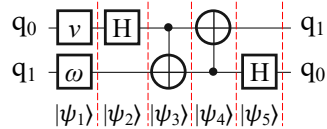

**Figure 4.** Five quantum states $|\psi\rangle$ of a $p$-SWAP gate for vertical-based operational gates. These $|\psi\rangle$ states illustrate the process of visually assigning the desirable $p$ for a set of swapped qubits ($q_0$ and $q_1$), using the BSA of two Bloch spheres (two XY-planes).

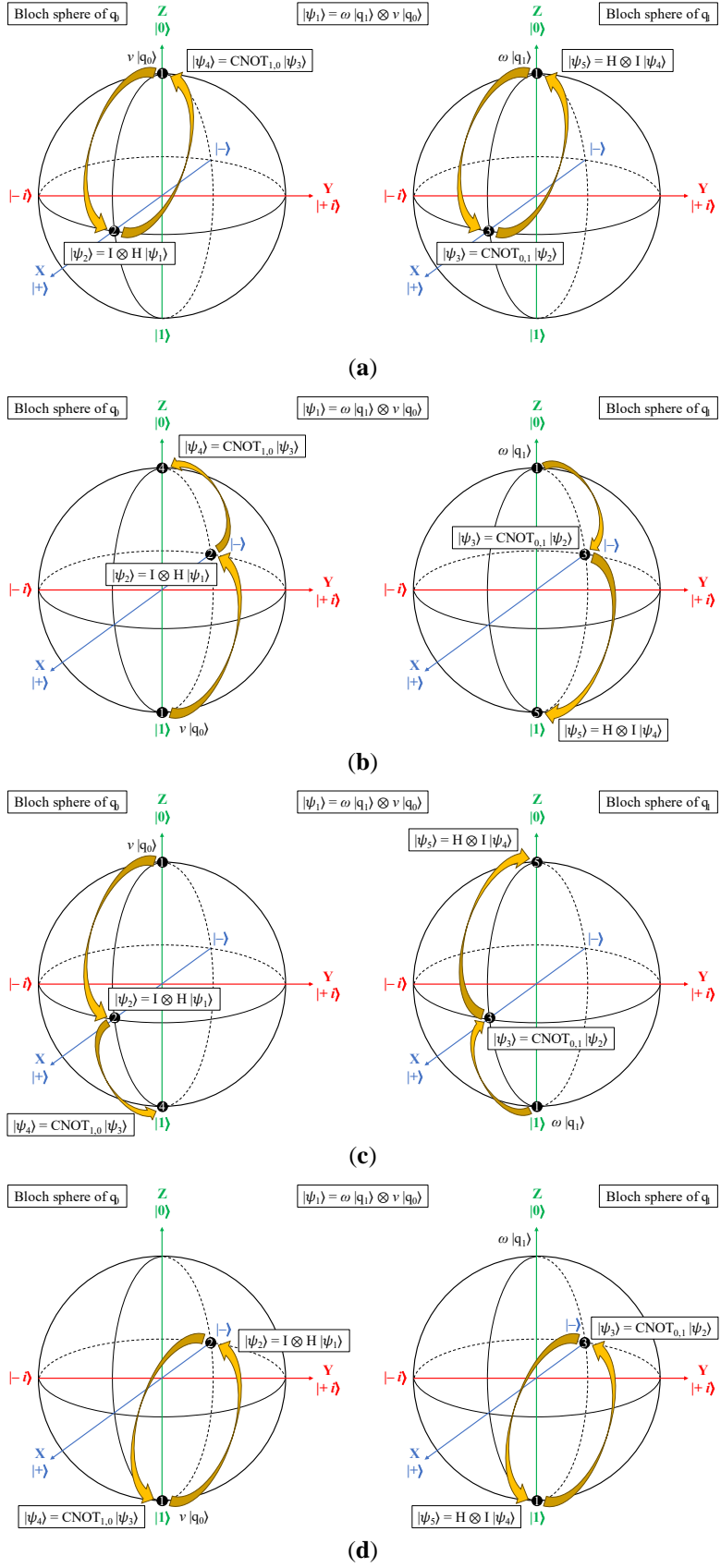

**Figure 5.** The quantum transitions (the yellow arrows) of the $|\psi\rangle$ states for the swapping outcomes of a $p$-SWAP gate using the BSA of two Bloch spheres (two XY-planes), for all permutative states of targets $|q_1 q_0\rangle$: (**a**) $|00\rangle \to |00\rangle$; (**b**) $|01\rangle \to |10\rangle$; (**c**) $|10\rangle \to |01\rangle$; (**d**) $|11\rangle \to |11\rangle$.

## 3. Results

Based on the six geometrical concepts stated in Table 1, different swapping outcomes of a *p*-SWAP gate are tested, as illustrated in Figure 6, for desirable *p* for a set of swapped qubits as follows.

1. The non-identical swapped qubits ($|01\rangle$ and $|10\rangle$) for $p = +\pi/2$ radians, as shown in Figure 6(a).
2. The identical swapped qubits ($|00\rangle$ and $|11\rangle$) for $p = +\pi/2$ radians, as shown in Figure 6(b).
3. The swapped qubits of $|00\rangle$ states for $p = +\pi$ radians, as shown in Figure 6(c).
4. The swapped qubits of $|01\rangle$ states for $p = +\pi$ radians, as shown in Figure 6(d).
5. The swapped qubits of $|10\rangle$ states for $p = +\pi$ radians, as shown in Figure 6(e).
6. The swapped qubits of $|11\rangle$ states for $p = +\pi$ radians, as shown in Figure 6(f).

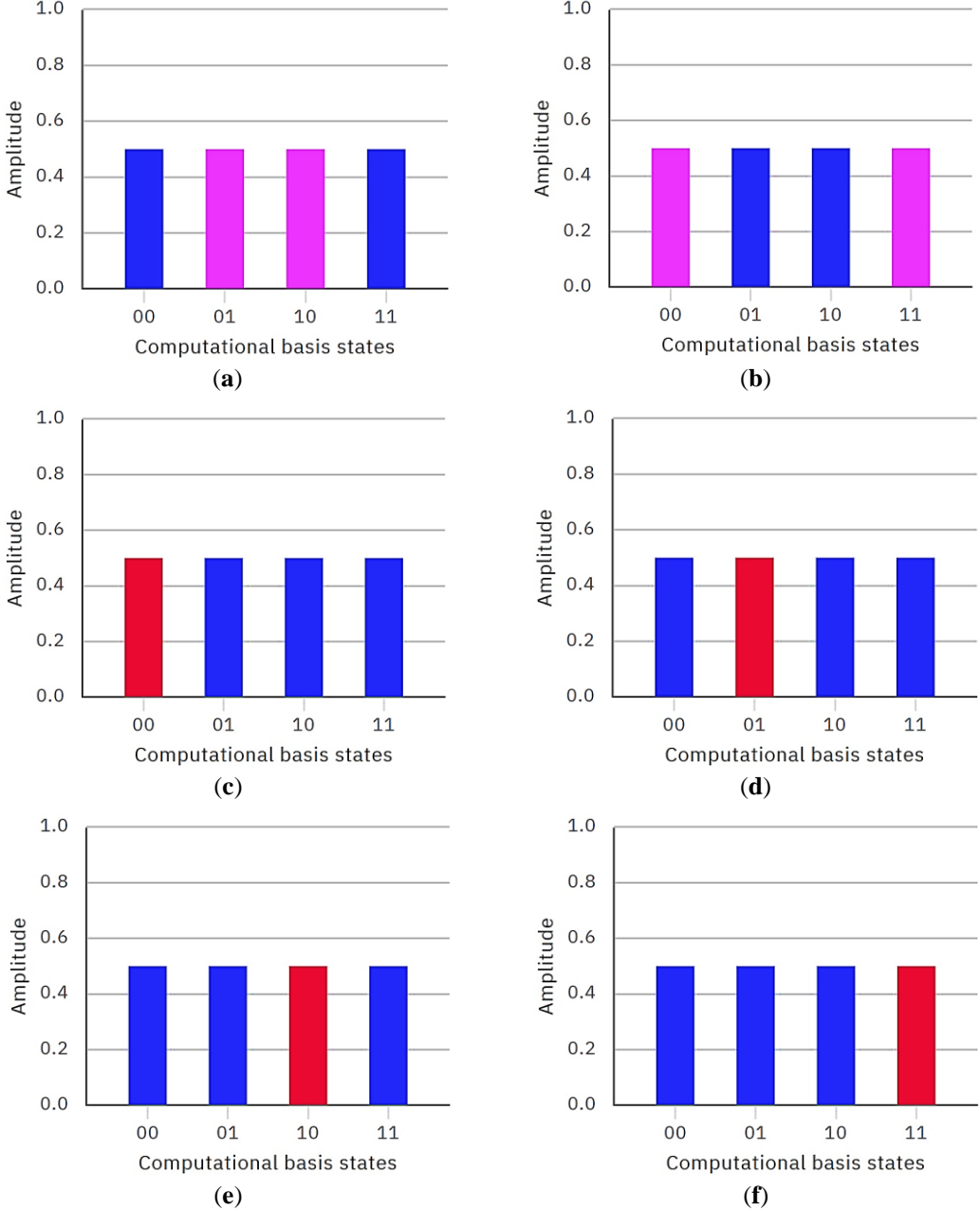

**Figure 6.** The Phase-based swapping outcomes of a *p*-SWAP gate for a desirable *p* for a set of swapped qubits: (**a**) the non-identical swapped qubits ($|01\rangle$ and $|10\rangle$); (**b**) the identical swapped qubits ($|00\rangle$ and $|11\rangle$); (**c**) the swapped qubits of $|00\rangle$ states; (**d**) the swapped qubits of $|01\rangle$ states; (**e**) the swapped qubits of $|10\rangle$ states; (**f**) the swapped qubits of $|11\rangle$ states, where the blue color indicates $p = 0$ radians, the pink color indicates $p = +\pi/2$ radians, and the red color indicates $p = +\pi$ radians.

In our research, the quantum circuits of the standard SWAP gate (see Figure 3(a)), the *i*SWAP gate (see Figure 3(b)), and our *p*-SWAP gate (see Figure 3(c)) are transpiled and evaluated using the `ibm_brisbane` quantum computer of 127 qubits [42,43]. These transpiled quantum circuits are evaluated in the context of our proposed "transpilation quantum cost (TQC)" [33,48]. The TQC is the sum of the total numbers of IBM 1-bit native $\sqrt{X}$, X, and RZ gates (denoted as $N_1$), IBM 2-bit native ECR gates (denoted as $N_2$), and the depth (denoted as D), where D is the critical longest path through all gates of $N_1$ and $N_2$. Figure 7(a) depicts the transpilation of one CNOT gate into an ECR gate using the `ibm_brisbane` quantum computer. Please observe that the technical specifications [28–30,49–51] of this quantum computer, e.g., the relaxation time (T1), decoherence time (T2), mean readout error of native gates, circuit layer operations per second (CLOPS), and quantum volume (QV), are not considered for the purpose of our research.

After transpilation, it was concluded that the transpiled quantum circuit of the *p*-SWAP gate has lower TQC and depth than those of the transpiled quantum circuit of the standard SWAP gate, for approximately 23% and 26% reduction of TQC and depth, respectively. However, the transpiled quantum circuit of the *p*-SWAP gate has identical TQC and depth to those of the transpiled quantum circuit of the *i*SWAP gate, because our introduced *p*-SWAP gate is the generalization concept of the *i*SWAP gate. Table 2 summarizes the calculated TQC for the final transpiled quantum circuits of the standard SWAP, *i*SWAP, and *p*-SWAP gates, as illustrated in Figure 7(b), Figure 7(c), and Figure 7(d), respectively.

**Table 2.** Summary of the transpilation quantum cost (TQC) for the final transpiled quantum circuits of the standard SWAP, *i*SWAP, and *p*-SWAP gates, using the `ibm_brisbane` quantum computer.

|  | $N_1$ | $N_2$ | D | TQC |
|---|---|---|---|---|
| Standard SWAP gate: | 20 | 3 | 15 | **38** |
| IBM *i*SWAP gate: | 16 | 2 | 11 | **29** |
| Our *p*-SWAP gate: | 16 | 2 | 11 | **29** |

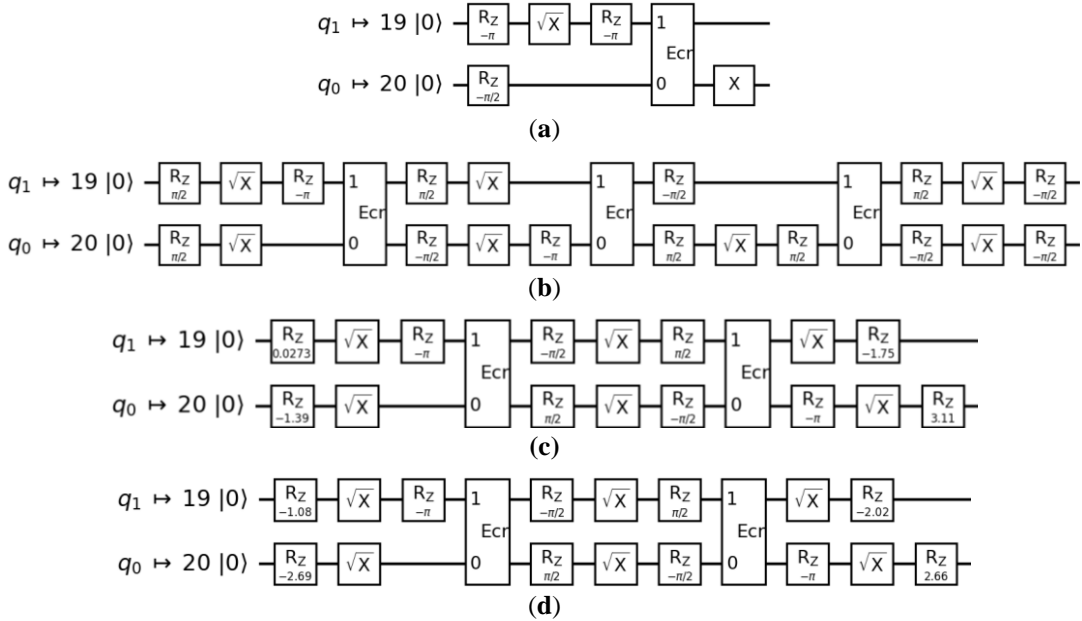

**Figure 7.** Final transpiled quantum circuits using `ibm_brisbane` quantum computer for: (**a**) one Feynman (CNOT) gate; (**b**) the standard SWAP gate consisting of three CNOT gates; (**c**) the *i*SWAP gate consisting of two CNOT gates; (**d**) our generic cost-effective *p*-SWAP gate consisting of two CNOT gates. Note that, for all IBM quantum computers of 127 qubits, the native CNOT gate has been replaced with the native ECR (echoed cross-resonance) gate [45–47].

Therefore, the *p*-SWAP gate can be employed as a generic and cost-effective transpilation library for the IBM quantum system, to replace the conventional transpilation of the standard SWAP gates, with the customizability of *p* based on the quantum operational purpose of the utilized Boolean and Phase oracles in various quantum applications and algorithms.

# 4. Conclusion

The standard quantum SWAP gate switches the indices of its two input qubits into two output qubits. This standard SWAP gate is constructed using three Feynman (CNOT) gates. For that, it is considered a cost-expensive quantum gate, when it is synthesized (transpiled) into a quantum computer. In this paper, we design a generic cost-effective swapping gate that is constructed using only two CNOT gates. Our swapping gate is termed the "$p$-SWAP gate", where $p$ is a customizable phase difference for $0 \leq p \leq \pm \pi$ radians. We introduce the $p$-SWAP gate as a generic cost-effective swapping framework, where (i) its generality comes from the customizability of $p$ for a set of swapped qubits for quantum Phase applications as well as for quantum Boolean applications, when $p$ is ignored, and (ii) its cost-effectiveness comes from the fewer utilized number of CNOT gates, as compared to the standard SWAP gate.

The quantum circuit of the $p$-SWAP gate is derived from the $i$SWAP gate proposed in [31], and the customizability of $p$ is geometrically chosen using our proposed Bloch sphere approach (BSA). In general, the BSA is a "geometrical design tool" for constructing cost-effective quantum gates, based on their rotational quantum operations in the projectional planes along the three axes of the Bloch sphere, without using the conventional design approach of matrices multiplication and unitary representations.

Experimentally, the quantum circuits of the standard SWAP and our $p$-SWAP gates are transpiled and evaluated using an IBM quantum computer, and it was concluded that the transpiled quantum circuit of the $p$-SWAP gate has lower quantum cost and depth than those of the transpiled quantum circuit of standard SWAP gate. Hence, the $p$-SWAP gate can be employed as a generic and cost-effective transpilation library, which can replace the conventional transpilation for the standard SWAP gates, with the customizability of $p$ that fulfills the quantum demands of Boolean and Phase applications.

**Data availability.** All relevant data are available from the authors upon request.

# References

[1] P. Kaye, R. Laflamme, and M. Mosca. *An Introduction to Quantum Computing*, Oxford University Press, New York, 2007.
[2] M.A. Nielsen and I.L. Chuang. *Quantum Computation and Quantum Information*, Cambridge University Press, Cambridge, 2010.
[3] R. LaPierre. *Introduction to Quantum Computing*, Springer, Cham, 2021.
[4] N. Earnest, C. Tornow, and D.J. Egger. "Pulse-Efficient Circuit Transpilation for Quantum Applications on Cross-Resonance-based Hardware", *Physical Review Research*, 3 (4), p. 043088, 2021.
[5] T. Hurant and D.D. Stancil. "Asymmetry of CNOT Gate Operation in Superconducting Transmon Quantum Processors Using Cross-Resonance Entangling", *arXiv:2009.01333*, 2020.
[6] A. Barenco, C.H. Bennett, R. Cleve, D.P DiVincenzo, N. Margolus, P. Shor, T. Sleator, J.A. Smolin, and H. Weinfurter. "Elementary Gates for Quantum Computation", *Physical Review A*, 52 (5), p. 3457, 1995.
[7] A. Al-Bayaty and M. Perkowski. "A Concept of Controlling Grover Diffusion Operator: A New Approach to Solve Arbitrary Boolean-based Problems", *Scientific Reports*, 14, pp. 1–16, 2024.
[8] A. Al-Bayaty and M. Perkowski. "BHT-QAOA: The Generalization of Quantum Approximate Optimization Algorithm to Solve Arbitrary Boolean Problems as Hamiltonians", *Entropy*, 26 (10), p. 843, 2024.
[9] P. Gokhale, T. Tomesh, M. Suchara, and F. Chong. "Faster and More Reliable Quantum Swaps via Native Gates". In *Proceedings of the 2024 International Conference on Parallel Architectures and Compilation Techniques*, pp. 351–362, 2024.
[10] S.F. Lin, S. Sussman, C. Duckering, P.S. Mundada, J.M. Baker, R.S. Kumar, A.A. Houck, and F.T. Chong. "Let Each Quantum Bit Choose Its Basis Gates". In *2022 55th IEEE/ACM International Symposium on Microarchitecture (MICRO)*, pp. 1042–1058, 2022.
[11] C. Duckering, J.M. Baker, A. Litteken, and F.T. Chong. "Orchestrated Trios: Compiling for Efficient Communication in Quantum Programs with 3-Qubit Gates". In *Proceedings of the 26th ACM International Conference on Architectural Support for Programming Languages and Operating Systems*, pp. 375–385, 2021.
[12] V. Gheorghiu, J. Huang, S.M. Li, M. Mosca, and P. Mukhopadhyay. "Reducing the CNOT Count for Clifford+T Circuits on NISQ Architectures", *IEEE Transactions on Computer-Aided Design of Integrated Circuits and Systems*, 42 (6), pp.1873–1884, 2022.
[13] A. Adedoyin, J. Ambrosiano, P. Anisimov, W. Casper, G. Chennupati, C. Coffrin, H. Djidjev, D. Gunter, S. Karra, N. Lemons, and S. Lin. "Quantum Algorithm Implementations for Beginners", *arXiv:1804.03719*, 2018.
[14] D. Camps, R. Van Beeumen, and C. Yang. "Quantum Fourier Transform Revisited", *Numerical Linear Algebra with Applications*, 28 (1), p. e2331, 2021.
[15] A. Mandviwalla, K. Ohshiro, and B. Ji. "Implementing Grover's Algorithm on the IBM Quantum Computers". In *2018 IEEE International Conference on Big Data*, pp. 2531–2537, 2018.

[16] J. Weidenfeller, L.C. Valor, J. Gacon, C. Tornow, L. Bello, S. Woerner, and D.J. Egger. "Scaling of the Quantum Approximate Optimization Algorithm on superconducting Qubit Based Hardware", *Quantum*, 6, p. 870, 2022.

[17] K. Blekos, D. Brand, A. Ceschini, C.H. Chou, R.H. Li, K. Pandya, and A. Summer. "A Review on Quantum Approximate Optimization Algorithm and Its Variants", *Physics Reports*, 1068, pp. 1–66, 2024.

[18] B. Tan and J. Cong. "Optimal Layout Synthesis for Quantum Computing". In *Proceedings of the 39th International Conference on Computer-Aided Design*, pp. 1–9, 2020.

[19] Y. Guowu, W.N.N. Hung, and S. Xiaoyu. "Majority-based Reversible Logic Gates", *Theoretical Computer Science*, 334 (13), p. 259274, 2005.

[20] M. Lukac, S. Nursultan, G. Krylov, and O. Keszöcze. "Geometric Refactoring of Quantum and Reversible Circuits: Quantum Layout". In *2020 23rd Euromicro Conference on Digital System Design* (*DSD*), pp. 428–435, 2020.

[21] M. Lukac, M. Perkowski, H. Goi, M. Pivtoraiko, C.H. Yu, K. Chung, H. Jeech, B.G. Kim, and Y.D. Kim. "Evolutionary Approach to Quantum and Reversible Circuits Synthesis", *Artificial Intelligence Review*, 20, pp. 361–417, 2003.

[22] K. Jang, G. Song, H. Kim, H. Kwon, H. Kim, and H. Seo. "Efficient Implementation of PRESENT and GIFT on Quantum Computers", *Applied Sciences*, 11 (11), p. 4776, 2021.

[23] S. Jaques, M. Naehrig, M. Roetteler, and F. Virdia. "Implementing Grover Oracles for Quantum Key Search on AES and LowMC". In *Advances in Cryptology–EUROCRYPT 2020: 39th Annual International Conference on the Theory and Applications of Cryptographic Techniques*, pp. 280–310, 2020.

[24] P. Fernández and M.A. Martin-Delgado. "Implementing the Grover Algorithm in Homomorphic Encryption Schemes", *Physical Review Research*, 6 (4), p. 043109, 2024.

[25] A. Bärtschi and S. Eidenbenz. "Deterministic Preparation of Dicke States". In *International Symposium on Fundamentals of Computation Theory*, pp. 126–139, 2019.

[26] C.S. Mukherjee, S. Maitra, V. Gaurav, and D. Roy. "Preparing Dicke States on a Quantum Computer", *IEEE Transactions on Quantum Engineering*, 1, pp. 1–17, 2020.

[27] S. Aktar, A. Bärtschi, A.H.A. Badawy, and S. Eidenbenz. "A Divide-and-Conquer Approach to Dicke State Preparation", *IEEE Transactions on Quantum Engineering*, 3, pp. 1–16, 2022.

[28] A. Mi, S. Deng, and J. Szefer. "Securing Reset Operations in NISQ Quantum Computers". In *Proc. of the 2022 ACM SIGSAC Conf. on Computer and Communications Security*, pp. 2279–2293, 2022.

[29] P. Murali, J.M. Baker, A. Javadi-Abhari, F.T. Chong, and M. Martonosi. "Noise-Adaptive Compiler Mappings for Noisy Intermediate-Scale Quantum Computers". In *Proc. of the 24th Int. Conf. on Architectural Support for Programming Languages and Operating Systems*, pp. 1015–1029, 2019.

[30] D. Koch, B. Martin, S. Patel, L. Wessing, and P.M. Alsing. "Demonstrating NISQ Era Challenges in Algorithm Design on IBM's 20 Qubit Quantum Computer", *AIP Advances*, 10 (9), 2020.

[31] IBM Quantum Documentation, "iSwapGate", Available at: https://docs.quantum.ibm.com/api/qiskit/qiskit.circuit.library.iSwapGate (Accessed on date: 11 December 2024).

[32] A. Al-Bayaty and M. Perkowski. "BSA: The Bloch Sphere Approach as a Geometrical Design Tool for Building Cost-Effective Quantum Gates", *Protocols.io*, 2024.

[33] A. Al-Bayaty and M. Perkowski. "GALA-n: Generic Architecture of Layout-Aware n-Bit Quantum Operators for Cost-Effective Realization on IBM Quantum Computers", *arXiv:2311.06760*, 2023.

[34] A. Al-Bayaty, "GALA-n Quantum Library", Available at: https://github.com/albayaty/gala_quantum_library (Accessed on date: 11 December 2024).

[35] A. Al-Bayaty, X. Song, and M. Perkowski. "CALA-n: A Quantum Library for Realizing Cost-Effective 2-, 3-, 4-, and 5-Bit Gates on IBM Quantum Computers using Bloch Sphere Approach, Clifford+T Gates, and Layouts", *arXiv:2408.01025*, 2024.

[36] A. Al-Bayaty, "CALA-n Quantum Library," Available at: https://github.com/albayaty/cala_quantum_library (Accessed on date: 11 December 2024).

[37] IBM Quantum Computing, "Qiskit Ecosystem", Available at: https://www.ibm.com/quantum/ecosystem (Accessed on date: 11 December 2024).

[38] X. Cheng, K.C. Chang, Z. Xie, M.C. Sarihan, Y.S. Lee, Y. Li, X. Xu, A.K. Vinod, S. Kocaman, M. Yu, and P.G.O. Lo. "A Chip-Scale Polarization-Spatial-Momentum Quantum SWAP Gate in Silicon Nanophotonics", *Nature Photonics*, 17 (8), pp. 656–665, 2023.

[39] M. Ni, R.L. Ma, Z.Z. Kong, X. Xue, S.K. Zhu, C. Wang, A.R. Li, N. Chu, W.Z. Liao, G. Cao, and G.L. Wang. "A SWAP Gate for Spin Qubits in Silicon", *arXiv:2310.06700*, 2023.

[40] J. Howard, A. Lidiak, C. Jameson, B. Basyildiz, K. Clark, T. Zhao, M. Bal, J. Long, D.P. Pappas, M. Singh, and Z. Gong. "Implementing Two-Qubit Gates at the Quantum Speed Limit", *Physical Review Research*, 5 (4), p. 043194, 2023.

[41] P. Niemann, L. Mueller, and R. Drechsler. "Combining SWAPs and Remote CNOT Gates for Quantum Circuit Transformation". In *2021 24th Euromicro Conference on Digital System Design (DSD)*, pp. 495–501, 2021.

[42] R.C. Farrell, M. Illa, A.N. Ciavarella, and M.J. Savage. "Scalable Circuits for Preparing Ground States on Digital Quantum Computers: The Schwinger Model Vacuum on 100 Qubits", *PRX Quantum*, 5 (2), p. 020315, 2024.

[43] IBM Quantum Platform, "Quantum Processing Units", Available at: https://quantum.ibm.com/services/resources?tab=systems&system=ibm_brisbane (Accessed on date: 11 December 2024).

[44] B. Schmitt and G. De Micheli. "Tweedledum: A Compiler Companion for Quantum Computing". In *Design, Automation & Test in Europe Conf. & Exhibition* (*DATE*), pp. 7–12, 2022.

[45] P. Jurcevic, A. Javadi-Abhari, L.S. Bishop, I. Lauer, D.F. Bogorin, M. Brink, L. Capelluto, O. Günlük, T. Itoko, N. Kanazawa, and A. Kandala. "Demonstration of Quantum Volume 64 on a Superconducting Quantum Computing System", *Quantum Science and Technology*, 6 (2), p. 025020, 2021.

[46] E. Pelofske, A. Bärtschi, and S. Eidenbenz. "Quantum Volume in Practice: What Users can Expect from NISQ Devices", *IEEE Trans. on Quantum Engineering*, 3, pp. 1–19, 2022.

[47] Y. Ji, K.F. Koenig, and I. Polian. "Optimizing Quantum Algorithms on Bipotent Architectures", *Physical Review A*, 108 (2), p. 022610, 2023.

[48] A. Al-Bayaty and M. Perkowski. "Cost-Effective Realization of n-Bit Toffoli Gates for IBM Quantum Computers Using the Bloch Sphere Approach and IBM Native Gates", *arXiv:2410.13104*, 2024.

[49] D.C. McKay, I. Hincks, E.J. Pritchett, M. Carroll, L.C. Govia, and S.T. Merkel. "Benchmarking Quantum Processor Performance at Scale", *arXiv:2311.05933*, 2023.

[50] A. Wack, H. Paik, A. Javadi-Abhari, P. Jurcevic, I. Faro, J.M. Gambetta, and B.R. Johnson. "Quality, Speed, and Scale: Three Key Attributes to Measure the Performance of Near-Term Quantum Computers", *arXiv:2110.14108*, 2021.

[51] A.W. Cross, L.S. Bishop, S. Sheldon, P.D. Nation, and J.M. Gambetta. "Validating Quantum Computers Using Randomized Model Circuits", *Physical Review A*, 100 (3), p. 032328, 2019.